# The Pathfinder Testbed: Exploring Techniques for Achieving Precision Radial Velocities in the Near-Infrared

Larry Ramsey*[a,b], Suvrath Mahadevan,[a,b], Stephen Redman[a], Chad Bender [a,b], Arpita Roy [a,b], Stephanie Zonak[a,], Steinn Sigurdsson[a,b,c], Alex Wolszczan [a,b,c]

[a]Dept. of Astronomy & Astrophysics, The Pennsylvania State University 525 Davey Laboratory, University Park, PA, USA 16802;   [b]Center for Exoplanets & Habitable Worlds, Penn State;
[c]Penn State Astrobiology Research Center, Penn State

## ABSTRACT

The Penn State Pathfinder is a prototype warm fiber-fed Echelle spectrograph with a Hawaii-1 NIR detector that has already demonstrated 7-10 m/s radial velocity precision on integrated sunlight. The Pathfinder testbed was initially setup for the Gemini PRVS design study to enable a systematic exploration of the challenges of achieving high radial velocity precision in the near-infrared, as well as to test possible solutions to these calibration challenges. The current version of the Pathfinder has an R3 echelle grating, and delivers a resolution of R~50,000 in the Y, J or H bands of the spectrum. We will discuss the on sky-performance of the Pathfinder during an engineering test run at the Hobby Eberly Telescope as well the results of velocity observations of M dwarfs. We will also discuss the unique calibration techniques we have explored, like Uranium-Neon hollow cathode lamps, notch filter, and modal noise mitigation to enable high precision radial velocity observation in the NIR. The Pathfinder is a prototype testbed precursor of a cooled high-resolution NIR spectrograph capable of high radial velocity precision and of finding low mass planets around mid-late M dwarfs.

## 1. INTRODUCTION

The Penn State Pathfinder instrument is a developmental spectrograph implemented to explore technical and data reduction issues involved with obtaining precision radial velocities (PRV) in the near infrared (NIR). The primary science driver the search for exoplanets around mid to late M dwarf stars. M dwarfs are the most numerous stars in the galaxy and are a very promising population to find earth mass planets in the Habitable Zone (Kasting and Catling[1]) around these stars. The low mass of M stars, less than 0.5 solar masses, together with close in orbits, 0.2 to 0.02 Astronomical Units (AU) yield radial velocity amplitudes for planets in the HZ around these stars that are well within current limits of ~1 m sec$^{-1}$. State of the art 1 m sec$^{-1}$ PRV and relies on precisely tracking and correcting for subtle changes in the instrumental spectral response function (SRF) and line position at the 10$^{-3}$ pixel level. This is done either by imposing an I$_2$ absorption spectrum on the target spectrum, which is limited to wavelengths in the visible spectrum (Butler et al[2]), or increasingly by using a simultaneous Th-Ar spectrum. The latter technique has been very successfully employed in ESO's HARPS (Pepe, et al.[3], Lovis et al.[4]). Mid to late M stars are very challenging for visible spectroscopy techniques and especially so for those using the I$_2$ reference technique. However, if one takes advantage of the fact that they emit most of their energy in the near infrared (NIR) hundreds of targets become accessible. Looking at the energy distribution mid to late M stars and it is clear that the 1 to 1.65 micron region is potentially a much better region studying PRV in these objects than the visible 0.5-0.7 μm region used by established techniques. Vacca and Rayner[5] used theoretical spectra to show there was sufficient radial velocity information in M star spectra and that the so called Q factor (Bouchy et al.[6]) in the NIR Y, J & H bands exceeded that in the visible (V & R band). The recent NSF/NASA commissioned ExoPlanet Task Force Report (Lunine et al.[7]) strongly recommends that ground based searches around M stars be 'fast-tracked' in its roadmap towards finding earth-like planets. Lessons learned from our systematic exploration with the Pathfinder instrument will be major stepping stones towards achieving 1-3m/s long term radial velocity precision and achieving this goal. In the last few years there has been significant progress in NIR PRV on the sun in the lab (Ramsey et al. [8]), on stars using a gas cell (Bean et al[9]) and atmospheric lines (Figueira et al.[10]).

*lwr@astro.psu.edu

We discuss here tests with Pathfinder at the Hobby-Eberly telescope (Ramsey et al[11]). The first version Pathfinder has already demonstrated ~7-10m/s precision on solar observations (Ramsey et al[8]) in the lab. Lessons learned from our systematic exploration of technical and data analysis issues with the Pathfinder will be a significant stepping stone towards achieving 1-3m/s long term radial velocity precision. and

## 2. PATHFINDER SYSTEMS

### 2.1 Spectrograph

The Pathfinder spectrograph as currently configured, schematically shown Figure 1, is similar to the one described in in Ramsey et al[8]. It is fiber fed warm pupil cross-dispersed echelle spectrograph that yields a resolution ($\lambda/\Delta\lambda$) of 50,000. It has been upgraded since to improve efficiency. The input fiber coupling was changed to match the f/6 collimator with the f/4.2 output of the fiber from the HET fiber instrument feed (Horner et al[4]). A pair of achromats optimized for the 1000nm region images the fiber from the FIF as well as one form the calibration system on a 0.1 mm slit. The diverging beam is folded on the way to the collimator which directs the light to a 71.5 degree blaze angle echelle grating. While undersized it is twice as efficient as that used in Ramsey et al[3]. This *R3* echelle operate in-plane at an off-Littrow angle of ~5$^o$ to achieve high dispersion. A 150 l/mm 5.46$^o$ blaze cross disperser optimized for the Y bandpass between 950 and 1100 nm gives the spectral format in Figure 5. We use the same camera system as described in detail in Ramsey et al.[8] which uses s parabolic mirror with a weak coma correcting lens right before the NIR dewar.

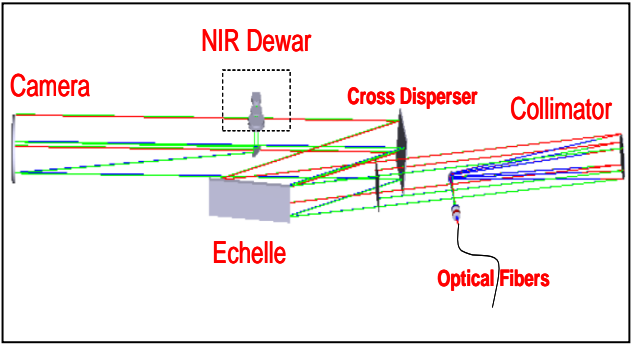

Figure 1: The schematic layout of the Pathfinder spectrograph.

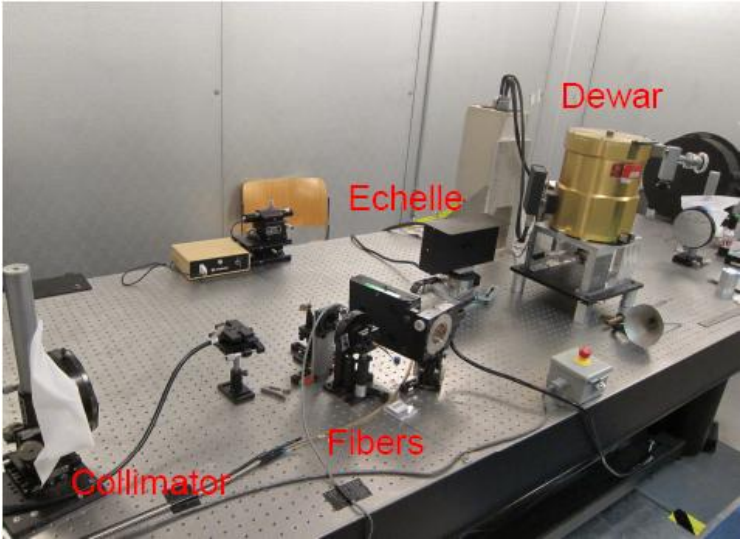

Figure 2: The Pathfinder on the optical bench inside a thermally insulating enclosure in the spectrograph room at HET.

## 2.2 Detector

The detector on Pathfinder is a Hawaii 1 1024x1024 NIR array with 0.018 mm pixels. This array is undersized to cover a full echelle free spectral range covering only ~33% of an order as illustrated in Figure 3. The next generation of Pathfinder will use a Teledyne H2RG and provide nearly full Y band coverage but will also require a new refractive camera.

Pathfinder is a warm pupil spectrograph and a thermal blocking filter is needed in the dewar to keep the detector from being swamped by the large thermal radiation background that at 2500 nm red end of its sensitivity. We accomplish this with a custom Y band filter and an additional commercial grade with thermal blocking filter in addition to a PK50 glass filter. With this setup the un-cooled instrument sees a background of ~3e/s/px, easily enabling 10 minute exposures with the Hobby Eberly Telescope.

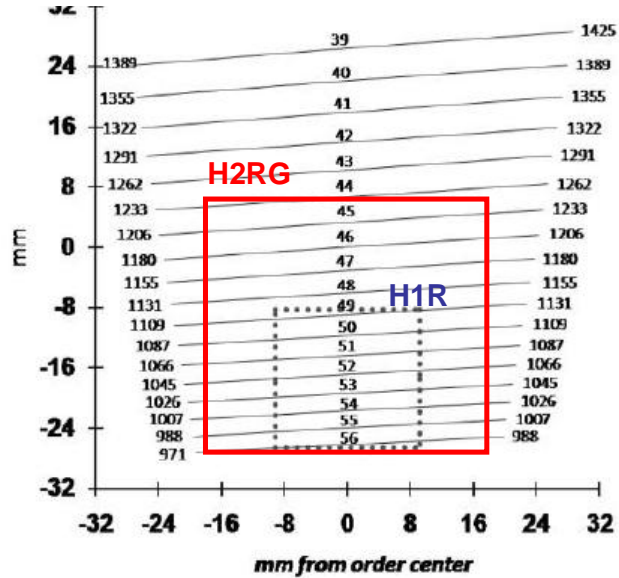

Figure 3: Illustrated is the free spectral range of the orders in the Y & J bands for our cross dispersed echelle configuration along with the coverage for our current detector, which is the same size as a H1R, and a H2RG

## 2.3 Calibration System

The major change in the calibration system from that described in Ramsey et al[8]. is that a U-Ne hollow cathode lamp is employed instead of a Th-Ar. Since Thorium-Argon lamps have weak Thorium lines and strong Argon lines (with are not stable at 1m/s) in the NIR we have used Pathfinder to explore Uranium hollow cathode lamps. These experiments make it clear that lamps with Neon as a filler gas have fewer bright lines in the NIR Y band, were we are currently focusing, than Argon which in turn leads to less scattered light and pixel saturation. Our experiments with Uranium (U) lamps convincingly show that U has a larger number of emission lines in the Y band, making it an attractive alternative to Thorium (See Figure 4). Uranium ($^{238}$U) fulfills many of the requirements needed of an element in a hollow cathode lamp for wavelength calibration: it is a heavy element (heavier than Thorium), has zero nuclear spin, and has a long half-life. It does have the disadvantage of a very small amount (~0.7%) of $^{235}$U in all naturally occurring Uranium, which may lead to very subtle isotope shifts between different lamps. A U-Ne hollow cathode lamp is now the primary calibration lamp for Pathfinder to enable accurate wavelength calibration as well as simultaneous monitoring of the instrument drift. We hope to extend our study of U-Ne to the J and H band to verifyU has more lines than Th in these bands as we expect.

A calibration fiber, fed with light from the U-Ne hollow cathode lamp is simultaneously fed to the spectrograph in parallel with the star fiber from the FIF. This simultaneous precision emission line reference enables monitoring of the instrument drift. A commercial paint mixer attached to the fibers enables the mitigation of modal noise effects (Baudrand and Walker[13]) which are significant in the NIR in our system. The appearance of the star plus calibration spectrum is shown in Figure 5.

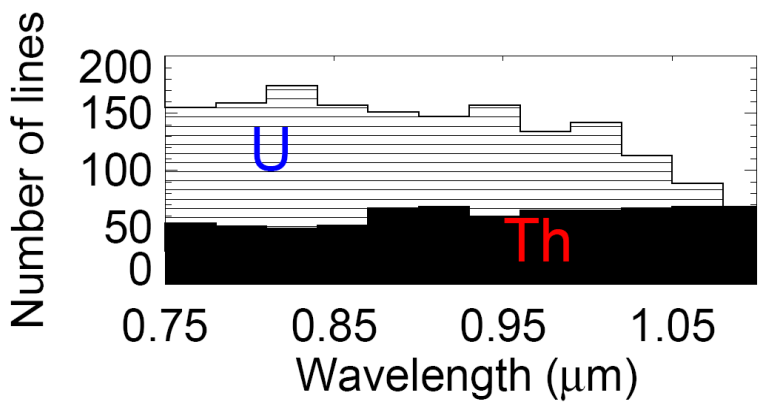

Figure 4: Uranium (U) and Thorium (Th) line densities are compared in the Y band

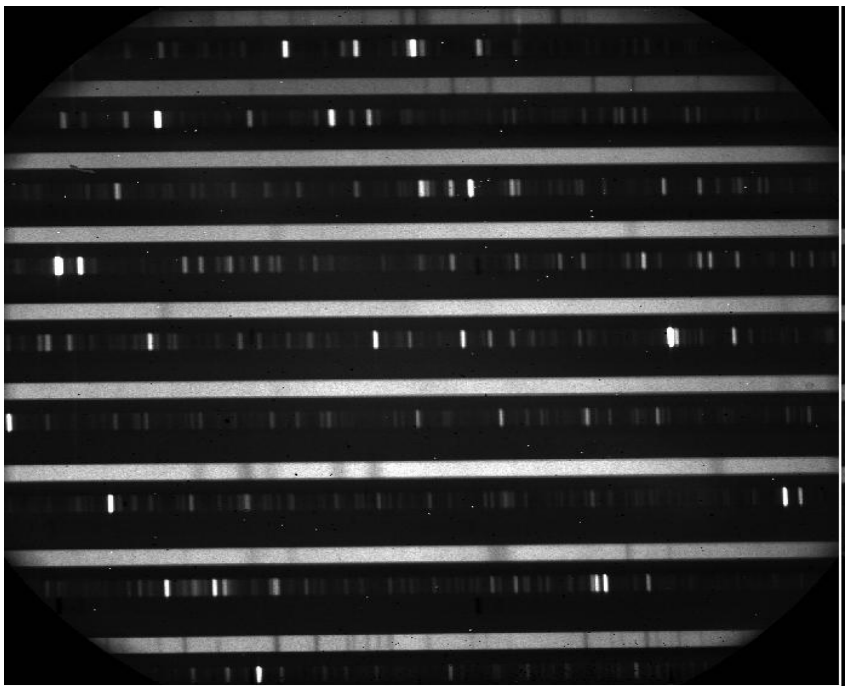

Figure 5: Star and Calibration Fiber, simultaneous exposure

## 3. PRELIMINARY RESULTS FROM HOBBY-EBERLY TELSCOPE

We installed Pathfinder at HET in March 2010 and conducted two test runs in April and May 2010. While pathfinder has achieved ~7 m/sec on the laboratory tests using a solar integrated light (Ramsey et al3), that was over hours and with a relatively large amplitude signal of ~300 m/sec. With the goal of 1-3 m/sec over months time scales it is critical to conduct telescope observation on real stellar targets. For this initial runs, we chose two relative bright relatively stable radial velocity targets. The first is the giant star HD106714 is known to be stable to ~10m/s. Binned velocities from 3 nights show a scatter of 11.3 m/s. Including the first two nights (where the instrument was still thermally stabilizing) and ~23m/s for all seven nights during the May run. The second target, Sigma Draconis, a know radial velocity stable star, was also observed. The binned velocities show a scatter of ~16.8 m/s including data from all nights and all observations in May. The HD106714 results are in Figure 6. These results are based on a preliminary analysis which we hope to refine in the coming months. We have additional data, including from our April 2010 run under analysis. All observations were taken in the same region in the Y band.

The fiber-fed Pathfinder instrument coupled to the HET is a very stable high resolution NIR spectrograph. Even in this testbed form we have demonstrated the ability to achieve 10-20m/s on stars. The next tests are planned for August 2010 where we will attempt to use a laser frequency comb developed at NIST. After that run we will return Pathfinder to Penn State for a planned upgrade to Pathfinder II incorporating what we have learned. We will substantially increase the efficiency and stability. A planned replacement of the current detector to a H2RG will significantly increase the wavelength coverage as well. Our goal with Pathfinder II is to attempt to achieve < 5m/s precision in the NIR in this next version which will be the final step before the implementation of a facility class instrument.

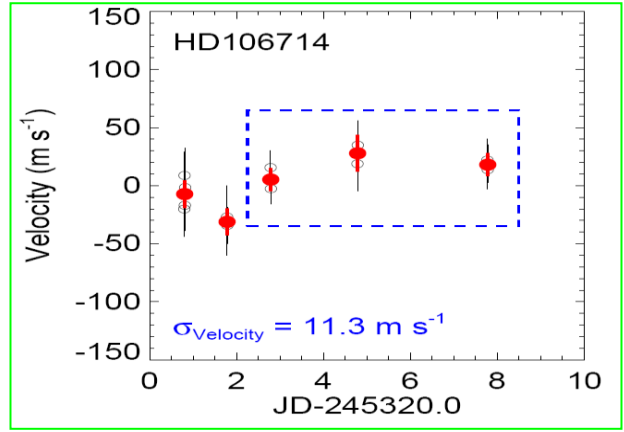

Figure 6: The velocity variation for HD106714 over the May 2010 test run .

## ACKNOWLEGMENTS

We acknowledge significant heritage from the Gemini PRVS proposal involving U Hertfordshire, UK ATC and Penn State. We are grateful to Hugh Jones for his input in this project. This work has been supported, in part by the Center for Exoplanets and Habitable Worlds and the Department of Astronomy & Astrophysics, Penn State. We also acknowledge support from NASA Origins grant NNX09AB34G and the NASA Astrobiology program via the Penn State Astrobiology Research Center.

## REFERENCES

[1] Kasting, J. F. and D. Catling, Evolution of a habitable planet, Ann. Rev. Astron. Astrophys. 41, 429-463 (2003).
[2] Butler, R. P.; Marcy, G. W.; Williams, E.; McCarthy, C.; Dosanjh, P.; Vogt, S. S. Publications of the Astronomical Society of the Pacific, v.108, p.500 (1996)
[3] Pepe, Francesco; Rupprecht, Gero; Avila, Gerardo; Balestra, Andrea; Bouchy, Francois; Cavadore, Cyril; Eckert, Wolfgang; Fleury, Michel; Gillotte, Alain; Gojak, Domingo; Guzman, Juan C.; Kohler, Dominique; Lizon, Jean-Luis; Mayor, Michel; Megevand, Denis; Queloz, Didier; Sosnowska, Danuta; Udry, Stephane; Weilenmann, Ueli in Instrument Design and Performance for Optical/Infrared Ground-based Telescopes. Edited by Iye, Masanori; Moorwood, Alan F. M. Proceedings of the SPIE, Volume 4841, pp. 1045-1056 (2003).
[4] Lovis, Christophe; Pepe, Francesco; Bouchy, François; Lo Curto, Gaspare; Mayor, Michel; Pasquini, Luca; Queloz, Didier; Rupprecht, Gero; Udry, Stéphane; Zucker, Shay in Ground-based and Airborne Instrumentation for Astronomy. Edited by McLean, Ian S.; Iye, Masanori. Proceedings of the SPIE, Volume 6269, 626. (2006).
[5] Vacca & Rayner unpublished research as part of PRVS proposal, (2006)_
[6] Bouchy, F., Pepe, F., & Queloz, D., "Fundamental photon noise limit to radial velocity measurements" A& A, 374, 733 – 739 (2001)
[7] Lunine, Jonathan I.; Fischer, Debra; Hammel, H. B.; Henning, Thomas; Hillenbrand, Lynne; Kasting, James; Laughlin, Greg; Macintosh, Bruce; Marley, Mark; Melnick, Gary; Monet, David; Noecker, Charley; Peale, Stan; Quirrenbach, Andreas; Seager, Sara; Winn, Joshua N., *Worlds Beyond: A Strategy for the Detection and*

*Characterization of Exoplanets Executive Summary of a Report of the ExoPlanet Task Force Astronomy and Astrophysics Advisory Committee Washington, DC June 23, 2008*, Astrobiology, Volume 8, Issue 5, pp. 875-88 (2008)

[8] Ramsey, L. W.; Barnes, J.; Redman, S. L.; Jones, H. R. A.; Wolszczan, A.; Bongiorno, S.; Engel, L.; Jenkins, J., *The Publications of the Astronomical Society of the Pacific, Volume 120, Issue 870, pp. 887-894* (2008).

[9] Bean, Jacob L.; Seifahrt, Andreas; Hartman, Henrik; Nilsson, Hampus; Wiedemann, Günter; Reiners, Ansgar; Dreizler, Stefan; Henry, Todd J., Ap.J. 713, p.410, (2010).

[10] Figueira, P.; Pepe, F.; Melo, C. H. F.; Santos, N. C.; Lovis, C.; Mayor, M.; Queloz, D.; Smette, A.; Udry, S., A&A, 511A, 55, (2010).

[11] L.W. Ramsey, M.T. Adams, T.G. Barnes, J.A. Booth, M.E. Cornell, J.R. Fowler, N.I. Gaffney, J.W. Glaspey, J. Good, P.W. Kelton, V.L. Krabbendam, L. Long, F.B. Ray, R.L. Ricklefs, J. Sage, T.A. Sebring, W.J. Spiesman, and M. Steiner, in Proceedings of the SPIE **3352**, *Advanced Technology Optical/IR Telescopes VI*, p.34. (1998)

[12] S. D. Horner, L. G. Engel, L. W. Ramsey, "Hobby Eberly Telescope medium-resolution spectrograph and fiber instrument feed", Proceedings of the SPIE **3355**, Part One, *Optical Astronomical Instrumentation*, p. 399. (1998).

[13] Baudrand, Jacques; Walker, Gordon A. H., *The Publications of the Astronomical Society of the Pacific, Volume 113, Issue 785, pp. 851-858. (2001).*